\documentclass[preprint,
amsmath,amssymb,aps,pre]{revtex4-2}

\usepackage{graphicx}
\usepackage{dcolumn}
\usepackage{bm}
 
\begin{document}

\preprint{APS/123-QED}

\title{Energy exchange in globally coupled mechanical phase oscillators} 

\author{Ra\'ul I. Sosa}
\email{iansosa996@gmail.com}
\author{Dami\'an H. Zanette}
\altaffiliation[Also at ]{Consejo Nacional de Investigaciones Cient\'{\i}ficas y T\'ecnicas (CONICET), Argentina}
\email{zanette@cab.cnea.gov.ar}
\affiliation{Centro At\'omico Bariloche and Instituto Balseiro, \\ Comisión Nacional de Energ\'{\i}a Atómica and Universidad Nacional de Cuyo, \\ 8400 San Carlos de Bariloche, R\'{\i}o Negro, Argentina}
 
\date{\today} 

\begin{abstract}
We study the stationary dynamics of energy exchange in an ensemble of phase oscillators, coupled through a mean-field mechanical interaction and added with friction and an external periodic excitation. The degree of entrainment between different parts of the ensemble and the external forcing determines three dynamical regimes, each of them characterized by specific rates of energy exchange. Using suitable approximations, we are able to obtain analytical expressions for those rates, which are in satisfactory agreement with results from numerical integration of the equations of motion. In some of the dynamical regimes, the rates of energy exchange show nontrivial dependence on the friction coefficients --in particular, non-monotonic behavior and sign switching. This suggests that, even in this kind of stylized model, power transfer between different parts of the ensemble and to the environment can be manipulated by a convenient choice of the individual oscillator parameters.
\end{abstract}

\maketitle

\section{Introduction}
\label{intro}
Coherent motion of mutually coupled oscillators is likely the most frequently invoked paradigm of self-organization in interacting dynamical systems. Different synchronization regimes, defined by various degrees of entrainment between the variables and/or the measurable properties that characterize oscillations, constitute a diversified assortment of archetypic forms of collective behavior, emerging from the combination of individual dynamics and interactions \cite{piko,nos}.  Foundational models of self-organized dynamics in ensembles of coupled oscillators were inspired by the observation of synchronization phenomena in biological populations \cite{winf}.  The quantitative representation of an oscillating element in terms of a phase measured along its cyclic trajectory, which underlies Kuramoto's celebrated phase-oscillator model \cite{kura}, has been profusely applied in diverse variations to the description of a broad class of chemical and biological systems, ranging from catalytic surface reactions \cite{cata}, to neural networks \cite{neur}, to ecosystems  \cite{ecol}.  

The first historically recorded scientific observation of a synchronization phenomenon involved mechanical oscillators (Christiaan Huygens' pendulum clocks, in 1665 \cite{piko}) but, in the context outlined in the preceding paragraph, the collective behavior of coupled dynamical systems governed by Newton equation has received relatively little attention. The joint dynamics of mutually interacting mechanical elements plays however a crucial role in many technological applications. A problem of particular interest regards the transfer of energy between the various parts of a given device, designed to perform specific functions by means of the power provided by a source.  This power is transformed and utilized by the device itself, and eventually delivered to other devices or to the environment \cite{vaka,dinc}. Especial regimes of power transfer, such as long-term circulation with low consumption, are desirable in devices such as pacemakers and certain sensors \cite{vanb,rein}. In contrast, fast dissipation becomes necessary when swift switching between different operational modes is required \cite{vaka,okam}. The interest in this phenomenology has recently been boosted by the discovery of unusual regimes of energy dissipation in nonlinear micro- and nanomechanical resonators, whose modal interactions are described by means of coupled mechanical oscillators \cite{polu,gutt,chen}. 

The Hamiltonian mean-field (HMF) model \cite{ruffo} is one of the few abstract systems formed by coupled mechanical oscillators whose collective behavior has been studied in detail, even in the thermodynamical limit. Formulated as a natural extension of Kuramoto's model for globally coupled phase oscillators to the realm of conservative dynamics, it has been used to analyze such phenomena as anomalous relaxation and inequivalence of statistical ensembles \cite{daux}. In this paper, we study the transfer of energy in an ensemble of phase oscillators globally coupled through the HMF model, to which we add energy input in the form of an external periodic force applied to one of the oscillators, and energy loss by friction. We see this system as an abstract representation of a generic mechanical device where energy is injected from outside, distributed among its internal components, and eventually given back to the environment. Our interest is focused on determining the rates of energy exchange between different parts of the ensemble, and how they depend on the individual parameters of each oscillator, specifically, on its damping coefficient. Underlying this interest is the possibility of controlling the flow of energy inside the system by choosing those parameters at an engineering stage. We begin by characterizing the main collective dynamical regimes that the ensemble can achieve in stationary motion. Next, we introduce a number of approximations that make it possible to obtain analytical solutions to the equations of motion. Finally, we calculate the corresponding rates of energy exchange and study their dependence on the relevant parameters, also comparing with results from numerical integration of the equations.          

\section{Mechanical phase oscillators with mean-field coupling} \label{sec2}

We consider a set of $N$ phase oscillators, each of them described by a dynamical variable $\theta_n (t) \in [0,2\pi)$ ($n=1,\dots, N$). The conservative part of the dynamics, which  encompasses global coupling between oscillators, is given by the HMF Hamiltonian \cite{ruffo,daux}
\begin{equation}
{\cal H} = \frac{1}{2}\sum_{n=1}^N p_n^2 -\frac{K}{2N}\sum_{n,m=1}^N \cos (\theta_m-\theta_n) ,
\end{equation}
where $p_n$ is the conjugate momentum associated with each $\theta_n$, and $K$ is the coupling strength. To this conservative dynamics, we add linear friction with positive damping coefficients $\gamma_n$, and a periodic external force of strength $F$ and frequency $\omega$ ($F,\omega>0$) applied to one of the oscillators ($n=1$, without generality loss) so that the resulting equations of motion read  
\begin{equation} \label{Newton}
\ddot \theta_n = \frac{K}{N} \sum_{m=1}^N \sin (\theta_m -\theta_n) -\gamma_n \dot \theta_n + F \sin (\omega t-\theta_1) \delta_{n1},
\end{equation}
where $\delta_{n1}$ is Kronecker's delta. The external force represents the interaction between the phase $\theta_1$ and an oscillator of prescribed phase $\omega t$, with the same functional form as the force due to coupling. Equation (\ref{Newton}) is of Tricomi's type \cite{tricomi}. A variant involving self-sustained phase oscillators, in the absence of external excitation, has recently been analyzed with emphasis on the stability of fixed-point and limit-cycle solutions \cite{gao}.

For ensembles of coupled phase oscillators, it is customary to define the mean-field (Kuramoto) order parameter  by averaging phases over the whole ensemble \cite{kura,nos}. In our case, because of the special role of oscillator 1, it is convenient to exclude it from the mean field, introducing instead
\begin{equation} \label{Z}
 R \exp (i \Theta) = \frac{1}{N-1} \sum_{n=2}^N \exp(i\theta_n).
\end{equation}
Moreover, in order to make the notation more compact, it is useful to change the time variable in Eq.~(\ref{Newton}) to $\tau = \omega t$ and rename $\omega^{-2} K \to K$, $\omega^{-2} F \to F$, and $\omega^{-1} \gamma_n \to \gamma_n$ for all $n$. These new parameters are all non-dimensional. With these definitions, the equations of motion become
\begin{equation} \label{mot1}
     \theta_1'' = \bar K R \sin (\Theta -\theta_1)-\gamma_1  \theta_1' +  F \sin ( \tau-\theta_1)
\end{equation}
for oscillator 1, and
\begin{equation} \label{motn}
 \theta_n'' = \bar K R \sin (\Theta -\theta_n)-\gamma_n \theta_n ' +  k \sin (\theta_1-\theta_n)
\end{equation}
for $n=2,\dots ,N$, where primes indicate differentiation with respect to $\tau$, and  $\bar K =(N-1) K/N$, $k=K/N$. For the sake of brevity, we call $\Omega$-set the sub-ensemble formed by  oscillators  $n=2,\dots ,N$.

As advanced in the Introduction, we are here interested in the mechanisms of energy exchange between the oscillators and with the environment. The mean energy per time unit transferred by all the acting forces to oscillator $n$, which equals the average change of its kinetic energy, is given by 
\begin{equation}
 w_n  =  \langle \dot \theta_n\ddot \theta_n \rangle_t = \omega^3  \langle \theta_n' \theta_n'' \rangle_\tau,
\end{equation}
where $\langle \cdot \rangle_t$ and $\langle \cdot \rangle_\tau$ indicate averages over the respective time variables. In accordance with the time rescaling applied above, we rename $\omega^{-3} w_n \to w_n$, which is now a non-dimensional quantity. Equations (\ref{mot1}) and (\ref{motn}) make it clear that $w_n$ has three well-differentiated contributions, namely,
\begin{equation} \label{rate}
     w_n  =  w_n^\Omega + w_n^\Gamma + w_n^F ,
\end{equation}
where $ w_n^\Omega = \bar K \langle R \sin (\Theta -\theta_n) \theta_n' \rangle_\tau $ is the power exchanged with the $\Omega$-set  and $ w_n^\Gamma  = -\gamma_n \langle ( \theta_n')^2 \rangle_\tau $ corresponds to the power dissipated by friction. Note that $ w_n^\Gamma$ is always negative, since it always measures a loss of energy. As for the third contribution, $ w_1^F =F \langle  \sin (\tau-\theta_1) \theta_1' \rangle_\tau$ is the power transmitted by the external force to oscillator 1. For $n=2,\dots, N$,  $ w_n^F  = k \langle  \sin (\theta_1-\theta_n)  \theta_n'\rangle_\tau$ stands for the power exchanged between these oscillators and oscillator 1, which mediates the transfer of energy from the external force to the $\Omega$-set.

\subsection{Dynamics of a single oscillator} \label{ssec2A}

Before proceeding to the study of the ensemble of coupled phase oscillators governed by Eqs.~(\ref{mot1}) and (\ref{motn}), it is useful to review the behavior of a single oscillator of the same kind, as it sheds light on the dynamical regimes observed for the ensemble. The  equation of motion for a single oscillator of phase $\theta(t)$ reads
\begin{equation} \label{motsingle}
     \theta ''= -\gamma  \theta ' +  F \sin (\tau-\theta).
\end{equation}
Taking as a new variable the phase shift between the external force and $\theta$, $\phi= \tau-\theta$, this equation becomes
\begin{equation} \label{motphi}
 \phi ''= -\gamma \phi ' +\gamma -  F \sin \phi \equiv -\gamma \phi '-\frac{dV}{d\phi}.
\end{equation}
This is an autonomous Newton equation for $\phi(t)$, subjected to friction with damping rate $\gamma$, and under the action of  a (non-periodic) potential $V(\phi) = - \gamma  \phi- F \cos \phi$. The combination of a constant-slope ramp and an oscillatory function in $V(\phi)$ determines two distinct regimes in the behavior of $\phi( \tau)$, as follows.  

If the slope (in modulus) is smaller than the force amplitude, $\gamma < F$, $V(\phi)$ has periodically distributed minima at the values of $\phi$ where $\sin \phi = \gamma /F$ and $\cos \phi >0$. Under the action of friction, $\phi (\tau)$ will eventually become trapped in the vicinity of one of the minima and asymptotically approach a fixed value. In this situation --which corresponds to a sufficiently strong external force and/or weak damping-- the oscillator becomes synchronized with the force, thus moving with the same frequency as the force at a fixed phase shift. 

If, on the other hand, the external force is too weak  and/or damping is too strong, i.e. $\gamma  > F$, the oscillator cannot ``follow'' the force. Under these conditions, the phase shift $\phi(\tau)$ grows indefinitely, exhibiting a precession-like motion controlled by the ramp and periodically modulated by the external force. 

In the limit $\phi''\to 0$, we get the overdamped version of Eq.~(\ref{motphi}),
\begin{equation} \label{motphi-ov}
 \phi '= 1 -  \frac{F}{\gamma} \sin \phi,
\end{equation}
which has been profusely discussed in the literature in connection with Kuramoto's phase oscillator model \cite{kura,piko,nos}. It shows an analogous separation between two dynamical regimes of synchronization and precession, at exactly the same critical point, $\gamma = F$. In the case of Eq.~(\ref{motphi-ov}), moreover, exact solutions can be found in terms of elementary functions. We show in the following that these two regimes reflect on the dynamics of the ensemble governed by Eqs.~(\ref{mot1}) and (\ref{motn}).  

\section{Three regimes of stationary collective motion} \label{sec3}

We have begun the study of Eqs.~(\ref{mot1}) and (\ref{motn}) by performing a preliminary numerical exploration of an ensemble of $N=100$ oscillators, in order to detect qualitatively different dynamical regimes in its long-time behavior. Numerical integration of the equations of motion was carried out by means of a standard scheme, as detailed in  Sect.~\ref{sec4}, for various combinations of the relevant parameters.

\begin{figure}[h]
\includegraphics[width= 0.9\columnwidth]{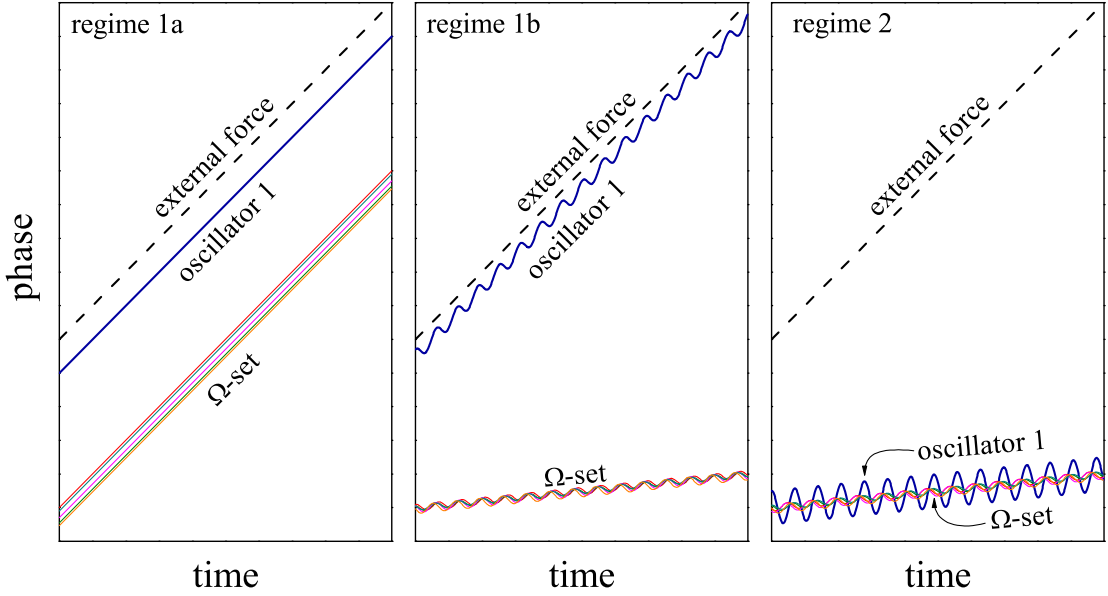} 
\caption{\label{fig1} Schematic representation of the long-time dependence of phases for the three regimes of oscillation entrainment analyzed in the text. The dashed line stands for the phase of the external force, which moves at constant frequency in all cases. The full bold line corresponds to oscillator 1. The thin lines in different shades stand for the phases of five selected oscillators in the $\Omega$-set.}
\end{figure}

It turns out that, depending on the parameter choice, the system exhibits three well-differentiated long-time stationary regimes, qualitatively characterized by diverse degrees of entrainment of oscillator 1 and the $\Omega$-set with each other, and with the external force. Figure \ref{fig1} shows a schematic representation of the long-time dependence of the phases of oscillator 1 (bold line) and a few oscillators in the $\Omega$-set (thin lines), as compared with that of the force (dashed straight line), which always grows linearly as time elapses. In the first two regimes, occurring for sufficiently large force amplitudes, oscillator 1 always ``follows'' the force. Depending on the strength of coupling, however, the $\Omega$-set can either synchronize to oscillator 1 (regime 1a, leftmost panel) or perform a non-synchronized, precession-like motion (regime 1b, central panel). In regime 1a, when coupling is large, the whole ensemble is fully synchronized with the force for long times, and all oscillators have constant phase shifts with respect to each other. In regime 1b, obtained for weak coupling, the oscillators in the $\Omega$-set move on the average with a much lower frequency, and their interaction with oscillator $1$ produces superposed small-amplitude oscillations. In turn, although oscillator 1 has on the average the same frequency as the external excitation,  its interaction with the $\Omega$-set also gives rise to oscillations around its mean motion. In regime 2, when the amplitude of the external force is small, neither oscillator 1 nor the $\Omega$-set can be entrained by the excitation. The whole ensemble moves on the average with a small frequency, and the effect of the force is limited to the appearance of superposed oscillations in the evolution of all phases.

As we show in Sect.~\ref{sec4} for the rates of energy exchange in the three regimes, numerical results are satisfactorily reproduced by an analytical approach to Eqs.~(\ref{mot1}) and (\ref{motn}) based on a few simplifying hypotheses. In the first place, we assume that the standard deviation of the damping coefficients $\gamma_n$, which we call $\sigma_\gamma$, is small as compared with the coefficients themselves. Numerically, we find that this assumption entails a small phase dispersion between the oscillators in the $\Omega$-set. On this basis, we conjecture  that the individual phases $\theta_n$ in the $\Omega$-set differ from their arithmetic average $\langle \theta \rangle=(N-1)^{-1} \sum_{n=2}^N \theta_n$ by   quantities $\epsilon_n =\theta_n - \langle \theta \rangle$ which are proportional to $\sigma_\gamma$. Neglecting terms of order $\sigma_\gamma^2$, the order parameter defined in Eq.~(\ref{Z}) becomes
\begin{equation}
    R \exp (i \Theta) = \exp (i \langle \theta \rangle)
\end{equation}
or, in other words, $R=1$ and $\Theta=\langle \theta \rangle$. Moreover, we assume that the average of the friction forces acting in the $\Omega$-set can be approximated as
\begin{equation}
\frac{1}{N-1} \sum_{n=2}^N \gamma_n \dot \theta_n = \langle \gamma \rangle \langle \dot \theta \rangle ,
\end{equation}
with $ \langle \gamma \rangle =(N-1)^{-1} \sum_{n=2}^N \gamma_n$ the mean damping rate. This assumption amounts to neglecting statistical correlations between individual velocities and damping coefficients.

Under the above hypotheses, Eq.~(\ref{mot1}) yields
\begin{equation} \label{m1a}
\theta_1'' = \bar K  \sin (\Theta -\theta_1)-\gamma_1 \theta_1 '+  F \sin (\tau-\theta_1)
\end{equation}
while, averaging Eq.~(\ref{motn}) over the $\Omega$-set, we get
\begin{equation} \label{mna}
    \Theta ''= -\langle \gamma \rangle \Theta '+ k \sin (\theta_1 - \Theta).
\end{equation}
These two equations constitute a self-consistent problem for the dynamics of the phase of oscillator 1 and of the average phase in the $\Omega$-set. By subtracting Eq.~(\ref{mna}) from Eq.~(\ref{motn}), moreover, we find the equation of motion for the individual deviations in the $\Omega$-set, $\epsilon_n$. Assuming $\bar K\gg k$, which holds for large $N$, we have
\begin{equation} \label{mea}
\epsilon_n ''=-\bar K  \epsilon_n -\gamma_n  \epsilon_n ' -(\gamma_n-\langle \gamma \rangle ) \Theta'.
\end{equation}
From this equation, we  see that in the case where all oscillators in the $\Omega$-set are identical --namely, when $\gamma_n=\langle \gamma \rangle$ for all $n$-- all the displacements $\epsilon_n$ tend to zero for long times. In this situation, the system asymptotically behaves as just two coupled oscillators excited by an external force, with phases $\theta_1$ and $\Theta$ governed by Eqs.~(\ref{m1a}) and (\ref{mna}).

In the following, we study the stationary (long-time) solutions of Eqs.~(\ref{m1a}) to (\ref{mea}) in the generic case in which the damping coefficients are not identical, within the three dynamical regimes pointed out above. Our treatment is based on considering different combinations of the parameters in suitably chosen limits, which are representative of each regime and, at the same time, allow for the derivation of analytical results.

\subsection{Regime 1a}
 
Both in regimes 1a and in 1b, oscillator 1 is entrained by the external force. This indicates that, in the right-hand side of Eq.~(\ref{m1a}), the last term is dominant over the first, which stands for the interaction with the $\Omega$-set. If, in addition, the external force dominates over the friction force given by the second term, we expect that the phase shift between oscillator 1 and the external force remains small for long times: 
\begin{equation} \label{theta1}
\theta_1 (\tau)=\tau - \eta (\tau),
\end{equation}
with $|\eta (\tau)| \ll 1$ for large $\tau$. The condition
\begin{equation} \label{reg1}
    F\gg \bar K , \gamma_1. 
\end{equation}
defines the approximation under which we deal with regimes 1a and 1b.

Equations (\ref{m1a}) and (\ref{mna}) yield, to the first significant order in $\eta$,
\begin{equation} \label{m1r1}
      \eta '' = \bar K  \sin (\tau-\Theta )-\gamma_1 ( \eta ' - 1) -  F \eta
\end{equation}
and
\begin{equation} \label{mnr1}
    \Theta ''= -\langle \gamma \rangle \Theta'+ k \sin (\tau- \Theta),
\end{equation}
respectively. In this last equation, we recognize the same equation of motion as for a single phase oscillator subjected to an external force, Eq.~(\ref{motsingle}). According to our discussion in Sect.~\ref{ssec2A}, the solutions describe either synchronization or precession of the average phase in the $\Omega$-set. The case of synchronization, for which
\begin{equation} \label{reg1a}
    \langle \gamma \rangle < k,
\end{equation}
defines the regime 1a considered here.

Given condition (\ref{reg1a}), the solution to Eq.~(\ref{mnr1}) asymptotically approaches the form $\Theta (\tau)= \tau- \Phi$, with $\sin \Phi =  \langle \gamma \rangle / k$ and $\cos \Phi >0$. Replacing this result into Eq.~(\ref{m1r1}) we get, for the stationary value of $\eta (\tau)$, 
\begin{equation} \label{eta1a}
    \eta =\frac{1}{F} \left[\gamma_1+(N-1) \langle \gamma \rangle \right]=\frac{\Gamma}{F},
\end{equation}
where $\Gamma=\sum_{n=1}^N \gamma_n$ is the sum of the damping coefficients all over the ensemble. Meanwhile, from Eq.~(\ref{mea}) we have
\begin{equation} \label{epsr1a}
    \epsilon_n = -\frac{1}{K} (\gamma_n-\langle \gamma \rangle),
\end{equation}
for long times. Note that, by virtue of Eqs.~(\ref{reg1}) and (\ref{reg1a}), the result obtained for $\eta$ is consistent with the assumption $|\eta (\tau)| \ll 1$, while Eq.~(\ref{epsr1a}) verifies the hypothesis that individual phase deviations in the $\Omega$-set are, on the average, proportional to the standard deviation of the damping coefficients, $\sigma_\gamma$.

In summary, under the present approximations, regime 1a corresponds to the situation where a sufficiently strong external force is able to fully entrain the oscillator ensemble. For long times, the whole system moves rigidly with the same frequency as the force. Different damping rates determine different individual asymptotic phases, so that the $\Omega$-set displays a phase dispersion proportional to the standard deviation of the $\gamma_n$.

\subsection{Regime 1b}

For $ \langle \gamma \rangle  > k$, as we have seen in Sect.~\ref{ssec2A}, the solution to Eq.~(\ref{mnr1}) for the average phase of the $\Omega$-set corresponds to a precession-like motion, not converging to a constant phase shift with oscillator 1 and the external force. Although this solution cannot be explicitly written down, the limit 
\begin{equation} \label{reg1b}
    \langle \gamma \rangle  \gg k
\end{equation}
admits an approximate analytical treatment whose output is in good agreement with numerical results (see Sect.~\ref{sec4}). Under condition (\ref{reg1b}), the large average friction implies that the $\Omega$-set moves at a small frequency, $\nu \ll 1$. Superposed to this motion, its phase  performs  oscillations of small amplitude, $A\ll 1$, with the same frequency as the external force, due to the effect of coupling with oscillator 1.  Meanwhile, since condition (\ref{reg1}) also holds in regime 1b, oscillator 1 should be close to the external force for long times. Due to the interaction with the $\Omega$-set, however, the phase shift $\eta (\tau)$ in  Eq.~(\ref{theta1}) is no more expected to asymptotically approach a constant value.

For  the average phase in the $\Omega$-set, we propose the {\it Ansatz}
\begin{equation} \label{anz1}
    \Theta (\tau) = A \cos [(1-\nu) \tau - \Psi]+\nu \tau.
\end{equation}
To the leading orders in  $\nu$ and $A$ and neglecting higher-harmonic contributions, replacement of this {\it Ansatz} into  Eq.~(\ref{mnr1}) gives 
\begin{equation} \label{A1b}
    A = \frac{k}{\sqrt{1+\langle \gamma \rangle^2}}
\end{equation}
for the  oscillation amplitude. Moreover, averaging the same equation over the fast oscillation of frequency $1-\nu$ in Eq.~(\ref{anz1}), we get
\begin{equation} \label{nu1b}
    \nu = \frac{1}{2}A^2 = \frac{k^2}{2(1+\langle \gamma \rangle^2)}
\end{equation}
for the precession frequency. The  phase $\Psi$ in Eq.~(\ref{anz1}) can also be obtained from this solution but, for brevity, it is not reported here. In fact, as is the case with the phases of all the oscillations considered in the following and in Sect.~\ref{ssec3C}, $\Psi$ is not involved in the calculation of the time-averaged rates of energy exchange considered in Sect.~\ref{sec4}. 

For the phase shift between oscillator 1 and the external force, Eq.~(\ref{m1r1}) yields   
\begin{equation} \label{etar1b}
    \eta (\tau) = \frac{\gamma_1}{F}+ \frac{\bar K  }{\sqrt{\left(F-1 \right)^2 +\gamma_1^2 }} \cos (\tau - \psi_1).
\end{equation}
Note that Eqs.~(\ref{A1b}) to (\ref{etar1b}), together with conditions (\ref{reg1}) and (\ref{reg1b}), are consistent with the assumptions $\eta, A, \nu \ll 1$. The long-time solution to Eq.~(\ref{mea}) is now
\begin{equation} \label{eps1b}
    \epsilon_n (\tau) = \frac{A (\gamma_n-\langle \gamma \rangle)}{\sqrt{\left(\bar K-1 \right)^2+\gamma_n^2}} \cos(\tau-\psi_n).
\end{equation}
for $n=2,\dots, N$, with $A$ given by Eq.~(\ref{A1b}).

In regime 1b, in summary, the external force is able to entrain oscillator 1, but the $\Omega$-set remains not entrained, and its average phase precedes with respect to that of the force. Due to its interaction with the $\Omega$-set,  oscillator 1 is not perfectly synchronized with the external force. Although, on the average, it moves with the same frequency as the force, the phase shift oscillates itself with a small amplitude. Likewise, the average phase in the $\Omega$-set oscillates around its predominantly preceding motion, and individual oscillators within the $\Omega$-set move with respect to each other with their own amplitudes and phases. 

\subsection{Regime 2} \label{ssec3C}

When external forcing is weak as compared to damping and coupling, oscillator 1 cannot be entrained by the force. In the limit
\begin{equation}
    F\ll \bar K , \gamma_1,
\end{equation}
it is observed that both oscillator 1 and the $\Omega$-set move, on the average, at  frequency $\nu \ll 1$, with a small phase difference between them. The marginal effect of the external force, however, makes them oscillate with small amplitude around their average precession motion, much as observed for the $\Omega$-set in regime 1b.  Therefore, our {\it Ansatz} for the time dependence of their phases  is  
\begin{equation}
\begin{array}{rl}
\theta_1(\tau) &= a_1 \cos[(1-\nu) \tau-\psi_1]+\nu \tau, \\   
\Theta(\tau) &= A \cos[(1-\nu) \tau-\Psi]+\nu \tau+\Theta_0. 
\end{array}
\end{equation}
Operating as for the $\Omega$-set oscillations in regime 1b, we find    
\begin{eqnarray}
 a_1 &=& \frac{F}{\sqrt{\left[\bar r\bar K-1+ r ( N-1)\right]^2+\left(\bar r \gamma_1 +r \Gamma \right)^2}} ,\\
  A &=& \frac{\sqrt{r}F}{\sqrt{\left[\bar r\bar K-1+ r ( N-1)\right]^2+\left(\bar r \gamma_1 +r \Gamma \right)^2}}, \label{A2}
\end{eqnarray}
for the oscillation amplitudes of phases $\theta_1$ and $\Theta$, and
\begin{equation} \label{nu2}
    \nu= \frac{1}{2} A^2 + \frac{\bar r \gamma_1}{2 \Gamma} a_1^2
\end{equation}
for the precession frequency. In the above equations, $\Gamma$ is defined as in Eq. (\ref{eta1a}), and
\begin{equation} \label{r}
    r= \frac{ k^2}{(k-1)^2 + \langle \gamma \rangle^2},  \ \ \ \bar r=1-r.
\end{equation}
Individual departures from the average phase $\Theta$ in the $\Omega$-set, given by $\epsilon_n (\tau)$  ($n=2,\dots, N$), have exactly the same form as in Eq.~(\ref{eps1b}), except that  $A$ is now given by Eq.~(\ref{A2}).

In regime 2, where both oscillator 1 and the $\Omega$-set precede with respect to the external force and move close to each other, the special dynamical status of oscillator 1 fades out --although it always mediates the effect of the external force on the system. Within our approximations, and depending on the value of the coefficient $r$ given by Eq.~(\ref{r}), the trajectory of oscillator 1 may lie completely inside the $\Omega$-set. From the viewpoint of its motion, therefore, oscillator 1 could remain qualitatively indiscernible from the rest of the ensemble. 
    
\section{Energy exchange in different regimes: Dependence on damping} \label{sec4}

In this section, we report the expressions for the different contributions to the rate of energy exchange in Eq.~(\ref{rate}), obtained from the analytical approximations discussed in Sect.~\ref{sec3} for the long-time dynamical regimes 1a, 1b, and 2. The contributions computed from these expressions are then compared with estimations deriving from  numerical integration of the equations of motion (\ref{mot1}) and (\ref{motn}), and their dependence with the damping rate of each oscillator is discussed. Numerical results were obtained for ensembles of $N=100$ oscillators,  using a standard fourth-order Runge-Kutta algorithm with an integration step two orders of magnitude lower than the shortest oscillation period in the system, which is determined by the frequency of the external force. Long-time conditions were achieved by letting transients which spanned several times the typical relaxation interval of energy damping. Time averages of energy exchange were computed over many oscillation periods in the long-time regime.      

In regime 1a, we find
\begin{eqnarray}
    w_1^\Omega &=&  -(N-1) \langle \gamma \rangle, \nonumber  \\
    w_1^\Gamma &=& -\gamma_1,  \label{w1a1} \\ 
    w_1^F &=&  (N-1)  \langle \gamma \rangle+\gamma_1 . \nonumber
\end{eqnarray}
for oscillator 1, and 
\begin{eqnarray}
    w_n^\Omega &=& \left(1-\frac{1}{N} \right) (\gamma_n-\langle \gamma \rangle),  \nonumber \\
    w_n^\Gamma &=& -\gamma_n, \label{w1a} \\ 
    w_n^F &=& \frac{1}{N} \gamma_n+ 
    \left(1-\frac{1}{N} \right) \langle \gamma \rangle, \nonumber
\end{eqnarray}
for $n=2,\dots,N$.  Note that $ w_n^\Omega + w_n^\Gamma + w_n^F =0$ for all $n$, as expected for any stationary dynamical state, where the incoming and outgoing mechanical power must compensate each other at each oscillator. As advanced in Sect.~\ref{sec2}, moreover, we have $w_n^\Gamma<0$, because damping always brings about a loss of energy.

\begin{figure}[h]
\includegraphics[width= 0.8\columnwidth]{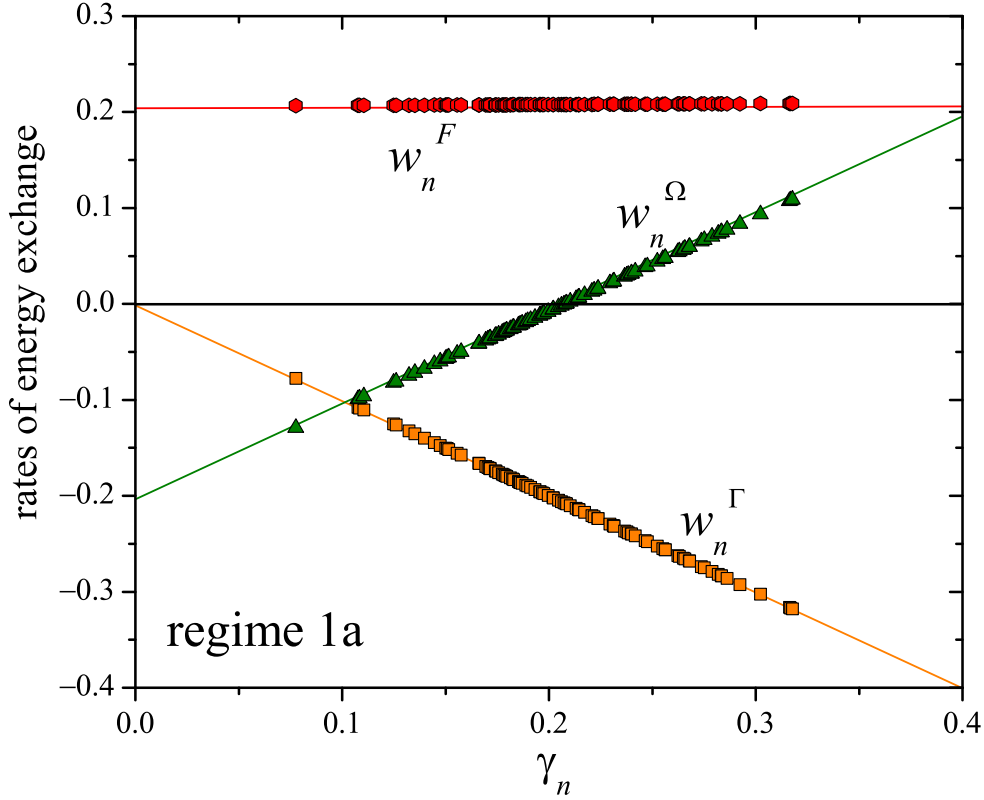} 
\caption{\label{fig2} Rates of energy exchange  --$w_n^\Omega$, $w_n^\Gamma$, and $w_n^F$-- in regime 1a, as functions of the damping coefficients of individual oscillators in the $\Omega$-set. Lines correspond to the analytical results of Eqs.~(\ref{w1a}), and symbols show numerical results for $N=100$, both for $\bar K=10$ and $F=100$. The damping coefficients were drawn from a normal distribution with mean $\langle \gamma \rangle =0.2$ and standard deviation $\sigma_\gamma =0.05$, yielding $\gamma_1=0.22$.}
\end{figure}

In this regime, the rates of energy exchange depend linearly on the damping coefficients. According to the first of  Eqs.~(\ref{w1a}), energy transfer between each oscillator in the $\Omega$-set and the $\Omega$-set itself occurs from the oscillators whose damping rate is smaller than the average towards those with larger coefficients. Meanwhile, the third of  Eqs.~(\ref{w1a}) shows that the energy transfer from the external force towards the $\Omega$-set, mediated by oscillator 1, is always positive. For large systems ($N\gg 1$), $w_n^F$ is virtually independent of $\gamma_n$, and it is solely determined by the average damping rate. Figure \ref{fig2} shows the three contributions for the oscillators in the $\Omega$-set as functions of the individual coefficients $\gamma_n$, with the parameters given in the caption. Lines and symbols correspond, respectively, to analytical and numerical results.  Their mutual agreement is very satisfactory.

In regime 1b,  the rates of energy transfer for oscillator 1 are
\begin{eqnarray}
    w_1^\Omega &=& -\frac{A^2}{2} (N-1) \langle \gamma \rangle ,\nonumber  \\
    w_1^\Gamma &=& -\gamma_1 ,  \label{w1b1}  \\ 
    w_1^F &=& \frac{A^2}{2} (N-1) \langle \gamma \rangle+ \gamma_1, \nonumber
\end{eqnarray}
while for the $\Omega$-set we have
\begin{eqnarray}
    w_n^\Omega &=& \frac{A^2 (\bar K-1) \bar K}{2} \frac{\gamma_n-\langle \gamma \rangle}{(\bar K-1)^2+\gamma_n^2},\nonumber  \\
    w_n^\Gamma &=& -\frac{A^2 }{2}\gamma_n  \frac{(\bar K-1)^2+ \langle \gamma \rangle^2}{(\bar K-1)^2+\gamma_n^2} ,  \label{w1b} \\ 
    w_n^F &=&  \frac{A^2}{2}   \frac{(\bar K-1) \bar K  \langle \gamma \rangle-\gamma_n ( \bar K-1-  \langle \gamma \rangle^2 )  }{(\bar K-1)^2+\gamma_n^2} , \nonumber
\end{eqnarray}
where the amplitude $A$ is given by Eq.~(\ref{A1b}). Again, $ w_n^\Omega + w_n^\Gamma + w_n^F =0$ and $w_n^\Gamma<0$ for all $n$. 

Since the amplitude $A$ does not depend on the individual damping coefficients, it is convenient to analyze the $\gamma_n$-dependence   of the rates of energy exchange in the $\Omega$-set through the normalized quantities $u_n^\Omega=2 A^{-2} w_n^\Omega$, $u_n^\Gamma=2 A^{-2} w_n^\Gamma$, and $u_n^F=2 A^{-2} w_n^F$. Besides $\gamma_n$, these quantities depend on the parameters $\bar K$ and $\langle\gamma\rangle$. As discussed for $w_n^\Omega$ in regime 1a, $u_n^\Omega$ switches its sign at $\gamma_n = \langle\gamma\rangle$, although now the change can be from positive to negative or vice versa. Moreover,  $u_n^F$ can also change sign, depending on the values of $\bar K$ and $\langle\gamma\rangle$. Careful inspection of Eqs.~(\ref{w1b}) reveals that there are three qualitatively different cases, with different sign combinations for $u_n^\Omega$ and $u_n^F$: $\bar K <1$ (case A), $1<\bar K<1+\langle\gamma\rangle^2$ (case B), and $1+\langle\gamma\rangle^2<\bar K$.

\begin{figure}[!h]
\includegraphics[width= 0.55\columnwidth]{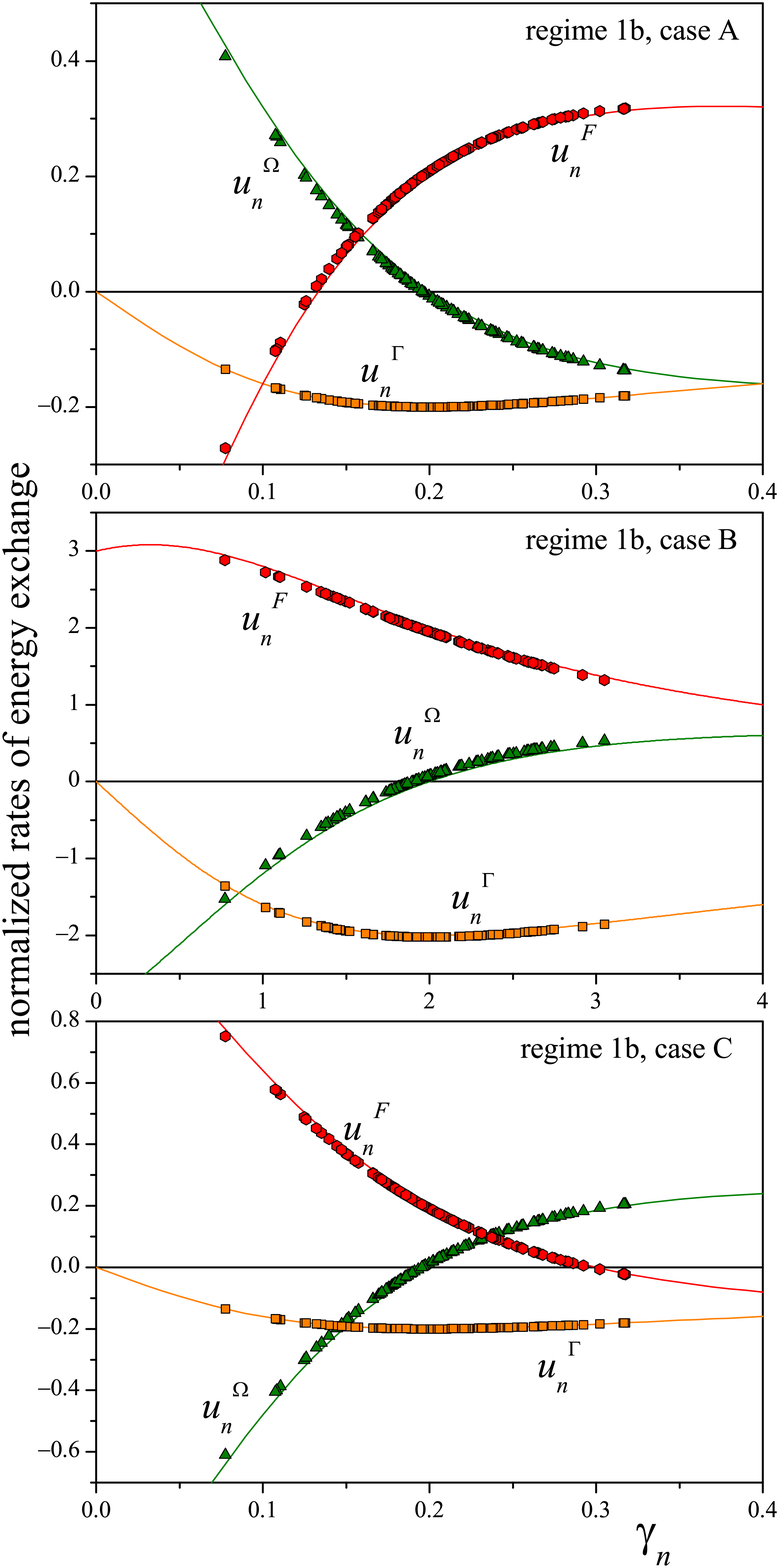} 
\caption{\label{fig3} Normalized rates of energy exchange  --$u_n^\Omega$, $u_n^\Gamma$, and $u_n^F$-- in regime 1b, as functions of the damping coefficients of individual oscillators in the $\Omega$-set. Lines and symbols respectively correspond to analytical, Eqs.~(\ref{w1b}), and numerical results for $N=100$, $F=100$. Upper panel, case A: $\bar K=0.8$, $\langle\gamma\rangle =0.2$. Middle panel, case B: $K=3$, $\langle\gamma\rangle =2$. Lower panel, case C: $K=1.2$, $\langle\gamma\rangle =0.2$. In cases A and C, the numerical values of $\gamma_n$ were the same as in Fig.~\ref{fig2}. In case B, the standard deviation was $\sigma_\gamma=0.5$, and $\gamma_1=1.6$.}
\end{figure}

Curves in Fig.~\ref{fig3} illustrate the normalized rates of energy exchange as functions of $\gamma_n$ in the three cases, as calculated from Eqs.~(\ref{w1b}) with the parameters specified in the caption.  In case A, as we see in the upper panel, $u_n^\Omega$ is positive for $\gamma_n < \langle\gamma \rangle$ and negative otherwise. Unlike what happens in  regime 1a (Fig.~\ref{fig2}) and in cases B and C, a net transfer of energy occurs in the $\Omega$-set from oscillators with large $\gamma_n$ to those with small $\gamma_n$. Meanwhile, $u_n^F$ switches from negative to positive values as $\gamma_n$ grows. This shows that, depending on $\gamma_n$, energy transfer between oscillator 1 and the $\Omega$-set can have either direction.  In case B (middle panel), $u_n^\Omega$ is negative for $\gamma_n < \langle\gamma \rangle$ and positive otherwise, while $u_n^F$ remains positive for any damping coefficient. In case C (lower panel) the changes of sign in $u_n^\Omega$ and $u_n^F$ are the opposite as in case A. 

Numerical results for the three cases, represented by symbols in  Fig.~\ref{fig3}, are again in very good agreement with the analytical approximation. Note that, according to Eqs.~(\ref{w1b}), the three rates of energy exchange decrease as $\gamma_n^{-1}$ for sufficiently large damping coefficients. For $u_n^\Gamma$, which vanishes for $\gamma_n=0$, this implies that it must reach a minimum for an intermediate value of the damping coefficient. This minimum is clearly visible in the three panels of Fig.~\ref{fig3}. Moreover,  since $u_n^\Omega$ crosses zero at $\gamma_n=\langle\gamma\rangle$, it must also attain a minimum or a maximum at a damping coefficient larger than $\langle\gamma\rangle$, and the same happens to $u_n^F$ in cases A and C. However, whether these extrema are effectively realized in a given oscillator ensemble depends on the dispersion of the damping coefficients. In the examples considered in Fig.~\ref{fig3}, the extrema in $u_n^\Omega$ and $u_n^F$ lie outside the interval spanned by the numerical values of $\gamma_n$. 

Finally, in regime 2 we find
\begin{eqnarray}
    w_1^\Omega &=& - (N-1) \langle \gamma \rangle \left( \frac{A^2}{2}  + \nu^2 \right),  \nonumber  \\
    w_1^\Gamma &=& -\gamma_1 \left( \frac{a_1^2 }{2}  + \nu^2 \right),  \label{w21} \\ 
    w_1^F &=& \frac{A^2}{2}  (N-1) \langle \gamma \rangle+\frac{a_1^2 }{2} \gamma_1 +\Gamma \nu^2 . \nonumber
\end{eqnarray}
for oscillator 1 and
\begin{eqnarray}
    w_n^\Omega &=& \frac{A^2 (\bar K-1) \bar K}{2} \frac{\gamma_n-\langle \gamma \rangle}{(\bar K-1)^2+\gamma_n^2}   
      + (\gamma_n-\langle \gamma \rangle) \nu^2,\nonumber \\ 
    w_n^\Gamma &=& -\frac{A^2 }{2}\gamma_n  \frac{(\bar K-1)^2+ \langle \gamma \rangle^2}{(\bar K-1)^2+\gamma_n^2}-\gamma_n \nu^2 ,  \label{w2} \\ 
    w_n^F &=&  \frac{A^2}{2}   \frac{(\bar K-1) \bar K  \langle \gamma \rangle-\gamma_n ( \bar K-1-  \langle \gamma \rangle^2 )  }{(\bar K-1)^2+\gamma_n^2}     +\langle \gamma \rangle \nu^2. \nonumber
\end{eqnarray}
for the $\Omega$-set, with $A$  given by  Eq.~(\ref{A2}). As in regimes 1a and b, $ w_n^\Omega + w_n^\Gamma + w_n^F =0$ and $w_n^\Gamma<0$ for all $n$. 

\begin{figure}[h]
\includegraphics[width= 0.55\columnwidth]{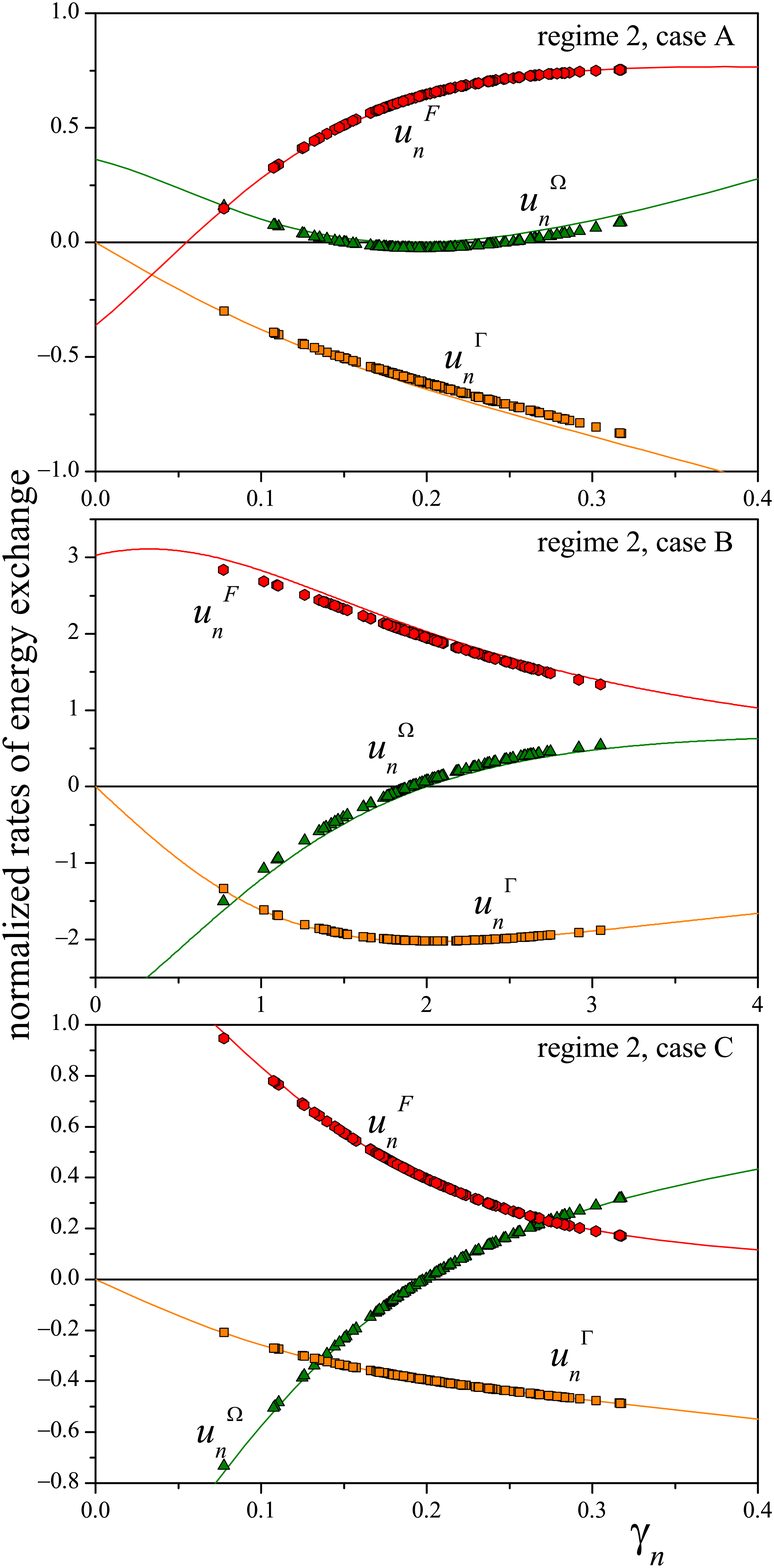} 
\caption{\label{fig4} As in Fig.~\ref{fig3} but for regime 2, with  $F=0.5$ and $\gamma_1=10$ in the three cases. }
\end{figure}

For the oscillators in the $\Omega$-set, the three rates of energy exchange consist of two contributions. The first one has exactly the same form as for regime 1b, Eqs.~(\ref{w1b}), although the amplitude $A$ has a different dependence on the parameters --cf.~Eq.~(\ref{A1b}). The second contribution is proportional to $\nu^2$ --cf.~Eq.~(\ref{nu2})-- and has a simple, at most linear, dependence on the individual damping coefficients. Since the frequency $\nu$ is small as compared with that of the external forcing, we expect that this contribution represents a moderate correction to the first term. Therefore, the normalized rates of energy exchange  are expected to roughly exhibit the same behavior as in regime 1b. The correction, however, may imply a quantitative difference if the distribution of damping coefficients is sufficiently broad. Figure \ref{fig4} shows analytical and numerical results in regime 2 within the same conditions as considered in Fig.~\ref{fig3} for regime 1b. For the three cases, the overall dependence on $\gamma_n$ is similar to that observed in regime 1b but, especially in cases A and C, deviations ascribable to  the linear dependence of the second contribution can be clearly perceived in the three normalized rates. 

The results presented in Figs.~\ref{fig2} to \ref{fig4} show that, depending on the parameters that control the system, a variety of qualitatively different situations occur in which concerns the contributions to energy transfer between oscillators and their dependence on the damping coefficients. In particular, the rates at which energy is exchanged through each of the mechanisms acting on the dynamics can vary non-monotonically with damping, and the direction in which energy is transferred between different parts of the system can also change.  

Regarding the rates of energy exchange associated to oscillator $1$ in the three regimes, Eqs.~(\ref{w1a1}), (\ref{w1b1}), and (\ref{w21}) make it clear that their dependence with the individual damping coefficient is much simpler than in the $\Omega$-set, so that they do not merit special discussion.  Let us just point out the approximate proportionality to $N$ in $w_1^\Omega$ and $w_1^F$, which can make these rates much larger than those corresponding to the other oscillators. This difference is directly related to the role of oscillator 1 as mediator of the power transferred from the external force.

\section{Conclusion}

We have studied the stationary dynamics of energy exchange in an ensemble of phase oscillators coupled through the Hamiltonian mean-field model, with added energy input in the form of an external periodic force applied to one of the oscillators, and energy loss by friction. For this system, we have shown the existence of three qualitatively disparate dynamical regimes, characterized by different degrees of synchronization with the external force. Using suitable approximations, we have obtained analytical expressions for the power transfer between different parts of the system, in each of the three regimes. The approximations involve limit assumptions on the relative weight of external excitation, coupling, and friction forces, as well as on the dispersion of the damping coefficients of individual oscillators. However, analytical solutions are in very good agreement with numerical results obtained under more relaxed conditions, which indicates that the range where each dynamical regime is observed must cover wider regions of parameter space. Nevertheless, the possibility of intermediate or mixed regimes cannot be discarded. 

In the regime where the whole system is fully entrained by the external excitation, all oscillators move rigidly with the same frequency as the force.  As a result of such simple dynamics, the power transfer shows monotonic dependence on the damping coefficients. In contrast, when the oscillator ensemble is only partially synchronized to the external force, this dependence is considerably richer, with non-monotonicity and changes of sign, and generally a broad diversity of behavior, depending on the  parameters of the system. This is an indication that power flow inside the ensemble can be manipulated by a convenient choice of the individual damping coefficients, making it possible to destine specific locations to distinct functions regarding the distribution and dynamics of energy --such as, for instance, points of maximum power intake or maximum energy dissipation. Such possibility is relevant to the design of devices ranging from sensors to switches \cite{vanb,rein,okam}.

A follow-up of this study may focus on the dynamics of energy exchange in an oscillator ensemble when the external excitation is turned off, and energy is freely left to dissipate by friction. Careful measurement of such ring-down stages, in fact, is a powerful tool in the detection of dynamical features such as nonlinearity, and currently plays a crucial role in the characterization of micro- and nanoscale mechanical devices  \cite{polu,gutt,chen,rd1,rd2,rd3}.

\bibliography{biblio} 

\end{document}